# On the self-consistency of DFT-1/2

Hanli Cui,[1,#] Shengxin Yang,[1,#] Kan-Hao Xue,[1,2,*] Jinhai Huang,[1] and Xiangshui Miao[1,2]

[1]School of Integrated Circuits, School of Optical and Electronic Information, Huazhong University of Science and Technology, Wuhan, 430074, China

[2]Hubei Yangtze Memory Laboratories, Wuhan 430205, China

#These authors contributed equally.

*Corresponding Author, Email: xkh@hust.edu.cn (K.-H. Xue)

# Abstract

DFT-1/2 is an efficient band gap rectification method for density functional theory (DFT) under local density approximation (LDA) or generalized gradient approximation. It was suggested that non-self-consistent DFT-1/2 should be used for highly ionic insulators like LiF, while self-consistent DFT-1/2 should still be used for other compounds. Nevertheless, there is no quantitative criterion prescribed for which implementation should work for an arbitrary insulator, which leads to severe ambiguity in this method. In this work we analyze the impact of self-consistency in DFT-1/2 and shell DFT-1/2 calculations in insulators or semiconductors with ionic bonds, covalent bonds and intermediate cases, and show that self-consistency is required even for highly ionic insulators for globally better electronic structure details. The self-energy correction renders electrons more localized around the anions in self-consistent LDA-1/2. The well-known delocalization error of LDA is rectified, but with strong overcorrection due to the presence of additional self-energy potential. However, in non-self-consistent LDA-1/2 calculations, the electron wavefunctions indicate that such localization is much more severe and beyond a reasonable range, because the strong Coulomb repulsion is not counted in the Hamiltonian. Another common drawback of non-self-consistent LDA-1/2 lies in that the ionicity of the bonding gets substantially enhanced, and the band gap can be enormously high in mixed ionic-covalent compounds like $TiO_2$. The impact of LDA-1/2-induced stress is also discussed comprehensively.

# I. INTRODUCTION

Density functional theory [1] (DFT) within the Kohn-Sham framework [2] has become a standard tool to calculate the electronic structure of solids, but with a frequently mentioned deficiency in recovering the fundamental band gap of semiconductors, at least under local density approximation [2,3] (LDA) or generalized gradient approximation [4,5] (GGA). Actually, the Kohn-Sham eigenvalues only describe a system with fixed ($N$) electron number, but the fundamental gap of a semiconductor involves the properties of three systems with $N$-1, $N$ and $N$+1 electrons, respectively. Hence, one should not resort to the origin LDA/GGA eigenvalues for the extraction of the fundamental gap. For molecules, the energy gap can be conveniently obtained by comparing the total energies of ($N$-1)-, $N$- and ($N$+1)-electron systems, while in a solid such $\Delta$-SCF (*i.e.*, computing the energy difference between two systems in terms of self-consistent-field runs) methods come with intricate technical details because stripping/adding only one electron from/to an infinite solid does not make sense. Hence, the $\Delta$-sol method as developed by Chan and Ceder [6], as well as another more recent approach by Ma and Wang [7], which does not involve empirical parameters, have been proposed as the proper $\Delta$-SCF-like methods for solids. Another proper approach is to use the quasi-particle concept and set up a connection between the fundamental gap of a semiconductor and quasi-particle excitation energies. The famous $GW$ approximation to the Hedin equations [8] is thus widely used in electronic structure calculations. Nevertheless, there is still a high practical demand in relating the Kohn-Sham eigenvalues to the quasi-particle band structure. In computational materials science and optoelectronics/microelectronics, it is usually expected that a one-shot self-consistent (SC) electronic structure run under the Kohn-Sham framework could yield the correct fundamental gap, detailed band structures as well as electronic density of states for a semiconductor [9,10]. This requires not resorting to the ($N$-1)- and ($N$+1)-electron systems, nor going beyond the Kohn-Sham framework (like introducing the quasi-particle equations).

To this purpose, there are still several solutions, where the most famous one is the hybrid functional approach. It is well-known that DFT performs in general better than Hartree-Fock because it considers the electron correlation effect, though it actually corresponds to the local, dynamical correlation in quantum chemistry [11]. However, the exchange term of DFT under LDA/GGA is incorrect, and it does not exactly cancel the spurious electron self-interaction. Introducing a part of the Hartree-Fock exchange thus should greatly improve the overall quality of exchange-correlation (XC) energy of LDA/GGA [12], better suppressing the electron self-interaction [13]. The consequence is that the convexity problem [14] of the XC in LDA/GGA will be greatly alleviated. In principle, the correct total energy should vary with respect to the electron number in a piecewisely linear manner (Koopmans' compliant) [15,16], provided that fractional occupation for orbitals is permitted. With hybrid functionals, the partial resolution of the convexity problem draws the functional closer to the Koopmans compliant. Given that the piecewise-linear property is fulfilled, it has been shown that derivative discontinuities will emerge at integral electron numbers [17], which adds to the fundamental gap of a semiconductor. The lack of derivative discontinuity in the XC of LDA/GGA has been regarded as the main reason of their band gap underestimation [18–22]. The self-interaction correction (SIC) method [23] is another well-known approach to rectify the exchange term and especially, the electron self-interaction. In addition, for strongly correlated materials such as CoO and NiO, the DFT+U method has been proposed [24], which is intrinsically orbital-dependent. None of these methods requires the self-energy operator or one-particle Green's function, and they carry out certain corrections to LDA/GGA in the reciprocal space.

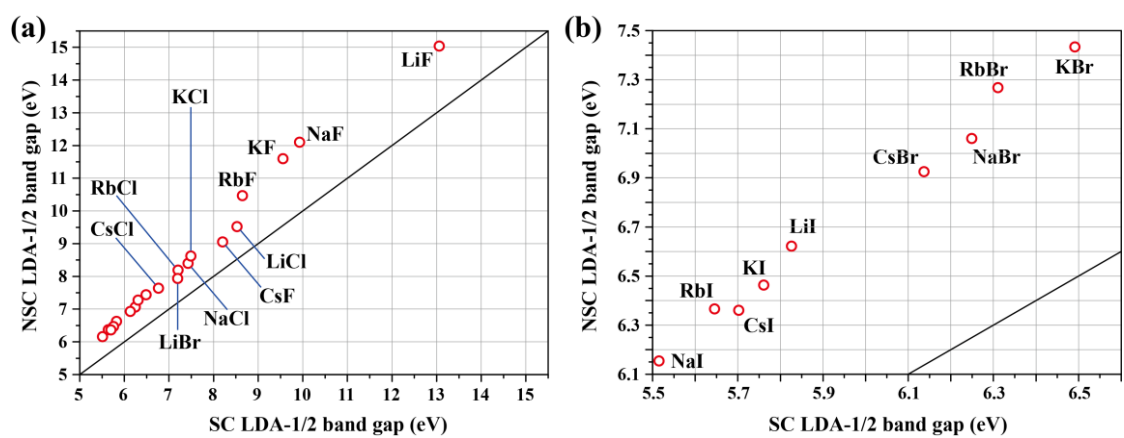

**Figure 1**. Comparison of calculated SC and NSC LDA-1/2 band gaps for alkaline halides. (a) A global comparison; (b) A magnified version for the lower gap regime. Data obtained from the calculations in this work.

The DFT-1/2 method [25–27], however, is a rather different approach for band gap rectification, in that it works in the real space and possesses almost the same efficiency as LDA/GGA [28,29]. This method stems from the Slater half-occupation technique [30], which was originally proposed for the Xα method [31]. The emergence of 1/2 roughly stems from the fact that the entity being excited is an entire electron, rather than an infinitesimal part of it as if electrons come only in their density [27]. In 2008, Ferreira, Marques and Teles generalized it into extended solids [25], proposing DFT-1/2, where DFT here represents either LDA or GGA. Several important new techniques have been introduced in that classic work. For instance, the hole in a semiconductor is assumed to possess non-null self-energy, which should be rectified in the excitation process. However, the conduction band (CB) electron is regarded as in a Bloch-like state with nearly null self-energy. Hence, the self-energy correction is only carried out for those atoms where the hole is localized. Secondly, the concept of self-energy potential (SEP) enables transferring the self-interaction information from an atomic calculation to a solid calculation, which greatly facilitates the self-energy correction as one-particle Green's function is no longer needed. Such transfer actually rectifies the band gap of a semiconductor in real space, in sharp contrast to SIC or DFT+U. Thirdly, an SEP cutoff radius ($r_{cut}$) is defined, which trims the SEP before its introduction to the solid, otherwise the long-range nature of Coulomb-type potential will render a total energy divergence [25,27,28]. The cutoff function for the SEP is therefore

$$\Theta(r) = \begin{cases} \left[1 - \left(\frac{r}{r_{cut}}\right)^p\right]^3 &, r \leq r_{cut} \\ 0 &, r > r_{cut} \end{cases} \quad (1)$$

where $p$ is a power index that is 8 by default.

While real space attachment of the SEP brings about great computational efficiency, it inevitably limits the application range of materials for DFT-1/2. For instance, the CB will also be pulled down provided that the CB electron has a certain probability to emerge in the spherical region of SEP. For covalent semiconductors, Xue *et al*. discovered that a shell-shape cutoff functional is much more effective in filtering out the hole, and it was shown that the generalized shell DFT-1/2 method [28] performs well for covalent semiconductors like Ge, GaAs, GaSb and InP that are technically important. The cutoff function becomes

$$\Theta(r) = \begin{cases} 0 & , r < r_{in} \\ \left\{1 - \left[\frac{2(r - r_{in})}{r_{out} - r_{in}} - 1\right]^p\right\}^3 & , r_{in} \leq r \leq r_{out} \\ 0 & , r > r_{out} \end{cases} \quad (2)$$

where $r_{in}$ and $r_{out}$ are the inner and outer cutoff radii, respectively, and $p$ (here $p$ must be an even integer) is suggested to be 20 to achieve a sharper trimming especially for the inner radius. Nevertheless, for other compounds such as $Li_2O_2$ [28] and $Cu_2O$ [32], which involve intermediate valency elements ($O^-$ and $Cu^+$), even shell DFT-1/2 fails to predict reasonable band gaps, because their electron and hole are entangled in the same shell-region. Therefore, DFT-1/2 (including shell DFT-1/2) could achieve high efficiency and accurate fundamental gap, provided that a carrier location analysis does not go against it like in $Li_2O_2$ or $Cu_2O$. More discussions on the DFT-1/2 and shell DFT-1/2 techniques are given in a recent review [27].

Nevertheless, DFT-1/2, including its variants like shell DFT-1/2 and DFT+A-1/2 [33], still suffers from some intrinsic issues on its theoretical side. For instance, three typical puzzles are summarized as below.

(i) In 2011, Ferreira and coworkers [26] pointed out that SC DFT-1/2 fails to predict sufficient band gap values for highly ionic insulators such as LiF and NaCl. They proposed that non-self-consistent (NSC) DFT-1/2, *i.e.*, sticking to the ground state electron density from the previous LDA/GGA calculation, should be used for highly ionic insulators instead of the conventional, standard SC DFT-1/2. The NSC DFT-1/2 band gaps are larger than SC DFT-1/2 band gaps, and examples of alkaline halides are illustrated in **Figure 1**. However, there is no definite prescription for when SC or NSC DFT-1/2 should be used. It was pointed out that when the hole is scattered in a series of valence bands (VBs), then SC DFT-1/2 must be the option. Yet, provided that the hole is only related to one VB, there is no absolute rule of choosing NSC or SC DFT-1/2. The issue of NSC DFT-1/2 causes a fatal problem to the uniqueness of the DFT-1/2 method. It is probably one of the most severe issues left for DFT-1/2.

(ii) In addition, there is a conceptual issue regarding the accuracy of SC DFT-1/2 in insulators. Since DFT-1/2 picks the VB hole from real space, band gap underestimation is still anticipated when the VB hole and CB electron are entangled in the same spatial locations.

Alkaline halides are actually closest to the ideal case of DFT-1/2, *i.e.*, the VB hole is mainly surrounding the anions, since the top of VB is dominated by halide anion states. It is therefore difficult to explain the DFT-1/2 gap underestimation in alkaline halides (*e.g.*, CsCl), considering that the DFT-1/2 band gap for some mixed ionic-covalent insulators are no less than experimental, such as anatase $TiO_2$. **Figures 2(a)-2(d)** compare the carrier spatial distributions of CsCl and anatase $TiO_2$, showing that the former is much closer to the ideal case of DFT-1/2. However, only the band gap of CsCl, rather than $TiO_2$, is underestimated through SC DFT-1/2, as evidenced by a comparison in **Figure 2(e)**. If one selects the standard LDA ground state charge density to set up the Hamiltonian, then the NSC LDA-1/2 band gap of CsCl, using the same $r_{cut}$ (3.2 bohr) as the SC DFT-1/2 case, is 8.20 eV and close to experimental (8.4 eV measured at 5 K [34]). If $r_{cut}$ is optimized again based on the LDA charge density, then the NSC LDA-1/2 band gap of CsCl increases to 8.44 eV at $r_{cut}$ = 2.7 bohr. For anatase $TiO_2$, SC LDA-1/2 yields a satisfactory band diagram (**Figure 2(f)**). The NSC LDA-1/2 band gap for anatase $TiO_2$, on the other hand, is much too large (7.49 eV compared with experimental value 3.38 eV [35], when using the same $r_{cut}$ of 2.4 bohr as in SC LDA-1/2; the individually optimized $r_{cut}$ for NSC LDA-1/2 calculation is 2.0 bohr, yielding a similar band gap of 7.85 eV). Such contradiction poses a challenge to the recognition of the ideal situation for SC DFT-1/2.

(iii) Another puzzle is related to the difference of shell DFT-1/2 compared with conventional DFT-1/2. It was shown that Ge is predicted to be a Γ-Γ direct gap semiconductor through LDA-1/2 calculation, but shell LDA-1/2 manages to recover its indirect band gap feature [28]. It is unclear why a shell-correction scheme not only increases the band gap value, but also transforms the type of that band gap.

Resolving these puzzles is important for the theoretical soundness and possible further development of DFT-1/2. To this end, the current research has been conducted, and the paper is organized as follows. In Sect. II, the self-consistency of DFT-1/2 is discussed within the Hohenberg-Kohn theorem. Subsequently, Sect. III presents SC and NSC calculations for highly ionic insulators. Sect. IV inspects the impact of self-consistency as well as SEP-induced stress on the energy band structures of covalent semiconductors. Sect. V demonstrates the surprising finding that NSC DFT-1/2 fails most significantly in intermediate cases, in between highly ionic insulators and typical covalent semiconductors. Sect. VI discusses the role of stress in DFT-1/2 calculations. Finally, the significance of self-consistency in DFT-1/2 and shell DFT-1/2 calculations is listed as the main conclusion of this work.

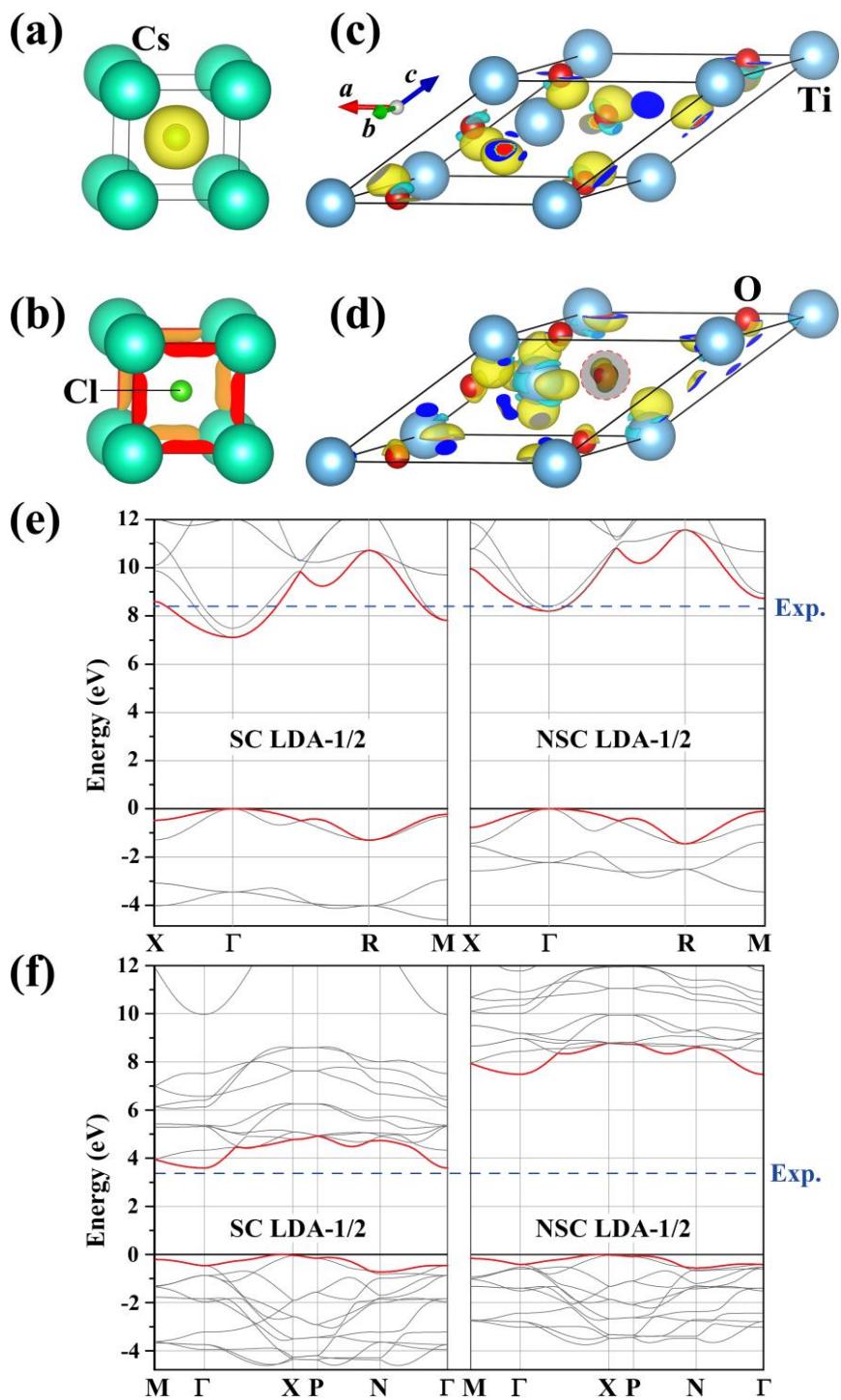

**Figure 2**. (a) Spatial distribution of VB hole in CsCl, with contour density 0.04 Å$^{-3}$ (same for below); (b) Spatial distribution of CB electron in CsCl; (c) Spatial distribution of VB hole in anatase TiO$_2$;

(d) Spatial distribution of CB electron in anatase TiO$_2$, where the shaded circle emphasizes its non-negligible probability to emerge near the O anions; (e) Band structure of CsCl calculated using SC and NSC LDA-1/2, with the same $r_{cut}$ = 2.7 bohr; (f) Band structure of anatase TiO$_2$ calculated using SC and NSC LDA-1/2, with the same $r_{cut}$ = 2.4 bohr.

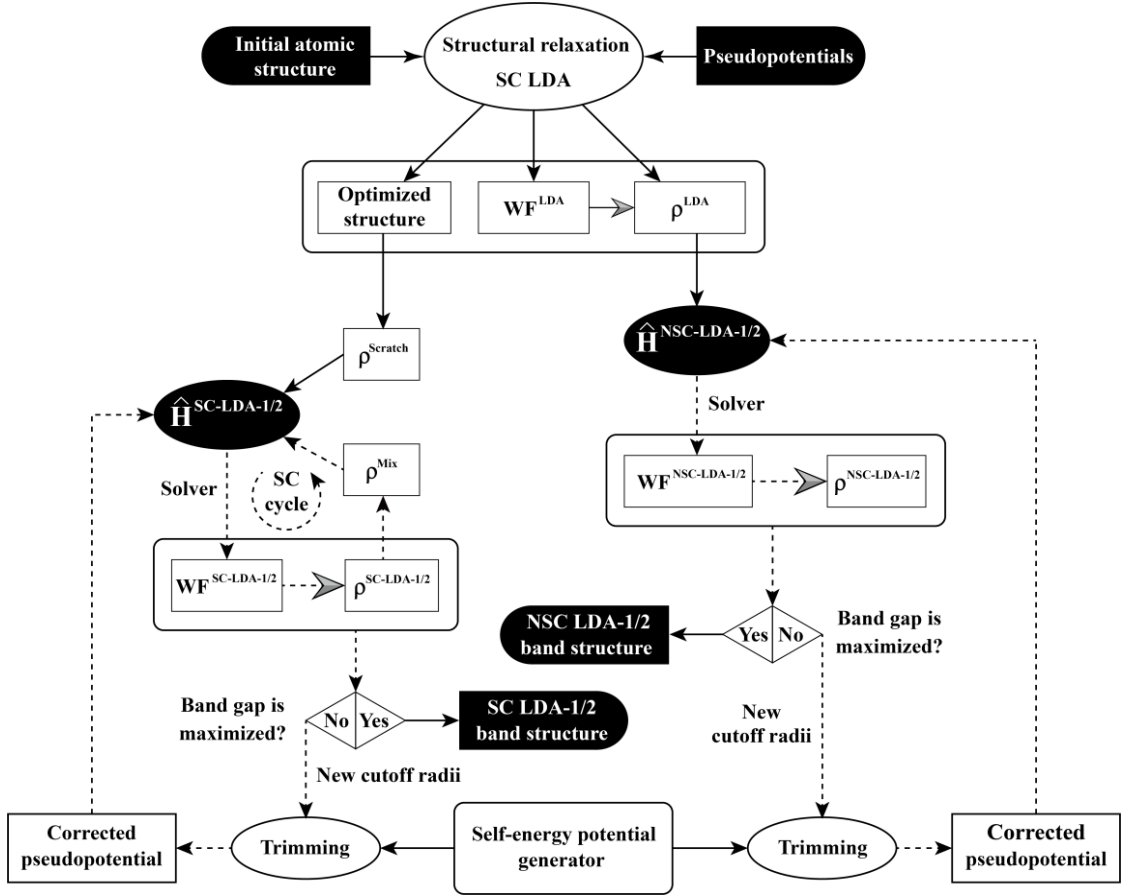

**Figure 3**. Procedure of an NSC LDA-1/2 calculation for a semiconductor/insulator, with comparison to an SC LDA-1/2 calculation. A dashed line means this step may be repeated multiple times in a cycle. WF stands for wavefunction and SC stands for self-consistent.

## II. THEORY AND COMPUTATIONAL SETTINGS

In SC DFT-1/2, the SEP is introduced into the solid as an additional external potential. The Hohenberg-Kohn theorem [1] states that the ground state charge density distribution is uniquely determined by the external potential. When one compares an SC LDA run with that of an SC LDA-1/2 run, the external potentials definitely differ. Hence, in principle an LDA-1/2 calculation should not be based on the corresponding LDA ground state charge density. The issue then becomes whether the new system, with different external potential and a distinct ground state charge density, resembles the prototype system in terms of electronic eigenvalues.

In NSC DFT-1/2 the situation is a bit more complicated, in that there is no relaxation allowed for the system to respond to the external potential, and the ground state charge is the same as that of LDA/GGA. The general procedures of NSC LDA-1/2 and SC LDA-1/2 are illustrated in **Figure 3**. In either case, the structure of the unit cell is first optimized using the LDA functional, and the LDA ground state charge density ($\rho^{LDA}$) is obtained. Subsequently, $\rho^{LDA}$ is utilized to construct the Hamiltonian of NSC LDA-1/2, serving as the charge density profile. The properly trimmed SEPs are also introduced into the unit cell, as part of the external potential for that Hamiltonian. The resulting Kohn-Sham equation is then solved, generating the electronic eigenvalues (*i.e.*, NSC LDA-1/2 band structure) and single-electron wavefunctions, which lead to a different charge density profile $\rho^{NSC-LDA-1/2} \neq \rho^{LDA}$. Such discrepancy lies at the heart of NSC LDA-1/2. In SC LDA-1/2, however, typically some initial guess ($\rho^{Scratch}$) for the charge density can be used to construct the initial Hamiltonian ($\rho^{LDA}$ may also be used as the initial guess if available), with also properly trimmed SEPs included. The Kohn-Sham equation corresponding to this initial Hamiltonian is solved to yield the new charge density $\rho^{[1]}$. Provided that $\rho^{[1]}$ is distinct from $\rho^{Scratch}$ by a non-negligible amount, mixing of a certain proportion of $\rho^{[1]}$ is done into $\rho^{Scratch}$ to yield a new charge density $\rho^{[2]}$. The Hamiltonian is re-constructed from $\rho^{[2]}$ and the SEPs, while the resulting Kohn-Sham equation is solved to yield $\rho^{[3]}$. Such self-consistent procedure continues until $\rho^{[i+1]}$ and $\rho^{[i]}$ well agrees through some convergence criterion (in practical it is usually the total energies rather than charge densities between two consecutive steps that are inspected for convergence), with the final charge density named $\rho^{SC-LDA-1/2}$. Hence, $\rho^{LDA}$, $\rho^{SC-LDA-1/2}$ and $\rho^{NSC-LDA-1/2}$ are all different.

The consequences of NSC DFT-1/2 are both a lack of orbital relaxation and a possible strong stress level in the calculation. On the one hand, the SEP pulls down the energy levels, thus the NSC nature forbids more electrons to be counted in the regions covered by SEP as far as the Hartree potential calculation is concerned, leading to a manual stress. This, however, cannot hinder the electron transfer viewed from the wavefunction. Indeed, even in NSC LDA-1/2 calculations, the true electron density distribution should still come from the electronic wavefunctions. The Hamiltonian utilizes some other charge density ($\rho^{LDA}$) for inter-electron repulsion, as if some unknown and unphysical screening effect exists. Therefore, whether more electrons enter or leave the SEP region should be judged from $\rho^{NSC\text{-}LDA\text{-}1/2}$ instead of $\rho^{LDA}$. This argument will be revisited later. On the other hand, the quality of NSC DFT-1/2 depends on the accuracy of the LDA/GGA ground state charge density. However, due to the inevitable approximations to the XC, the LDA/GGA ground state charge density still differs from the exact ground state charge density. While it is difficult (though not impossible, since experimentally it may be feasible to measure the charge density) to obtain the exact ground state charge density, more advanced techniques such as hybrid functionals may serve as a benchmark for the charge density, as the spurious electron self-interaction is supposed to be greatly suppressed in hybrid functional calculations.

In this work, we adopt the range-separated HSE06 hybrid functional [36,37] as a benchmark for the DFT-1/2 ground state charge density. It is shown that the LDA or GGA ground state charge density is different from the HSE06 charge density, to a non-negligible extent. The SC DFT-1/2 charge density ($\rho^{SC\text{-}DFT\text{-}1/2}$) goes towards the HSE06 charge direction, but it over-corrects to screen the external potential. The accurate NSC DFT-1/2 band gaps for alkaline halides do not accompany satisfactory electronic structure details. It is shown that SC DFT-1/2 is indispensable even in highly ionic insulators. Moreover, the impact of stress on SC DFT-1/2 calculations for covalent semiconductors will be discussed.

To eliminate any role of the exchange enhancement effect, besides HSE06 reference calculations, we have adopted LDA for the XC energy, within the fitted form of Perdew-Zunger [23] based on the quantum Monte Carlo results on uniform electron gas by Ceperley and Alder [38]. The projector augmented-wave method [39,40] was used, within the Vienna *Ab initio* Simulation Package [41,42] (VASP 5.4.4). Pseudopotential information and detailed computational settings are given in the **Appendix**.

# III. SELF-CONSISTENCY OF DFT-1/2 IN HIGHLY IONIC INSULATORS

In a seminal work of 2011 [26], Ferreira, Marques and Teles stated that NSC DFT-1/2 calculation should be performed for highly ionic insulators such as those alkaline halides. The reason lies in that the hole is localized exact on an anion, such as on F in the case of LiF, similar to the atomic half-occupation calculation. Nevertheless, in NSC DFT-1/2 calculations for solids, there is a lack of correlation effect against the external SEP. To evaluate whether NSC DFT-1/2 is a proper choice, we select two representative alkaline halides LiF and CsCl, with distinct crystalline structures, *i.e.*, the rock salt structure and the CsCl structure. Their LDA as well as SC LDA-1/2 charge densities were calculated, and compared with the HSE06 charge densities by subtraction. To this end, our HSE06 calculation was still based on the LDA-optimized lattice structures. As shown in **Figure 4**, $\rho^{LDA}$ has less amount of electron located in the vicinity of F anions, compared with the reference HSE06 result ($\rho^{HSE}$). This is reasonable because of the well-known delocalization error of LDA [43], spreading out the valence electrons. On the other hand, $\rho^{SC-LDA-1/2}$ shows more electrons located in the vicinity of the F anion, compared with that of HSE06, as illustrated in **Figure 4(b)**. Hence, LDA-1/2 rectifies the delocalization error towards the correct direction, but it over-corrects to a great extent because the external SEP attracts more electron towards the vicinity of the F anion. The SC LDA-1/2 band gap of LiF is 13.06 eV ($r_{cut}$ = 2.1 bohr), slightly lower than experimental value 14.2 eV [44].

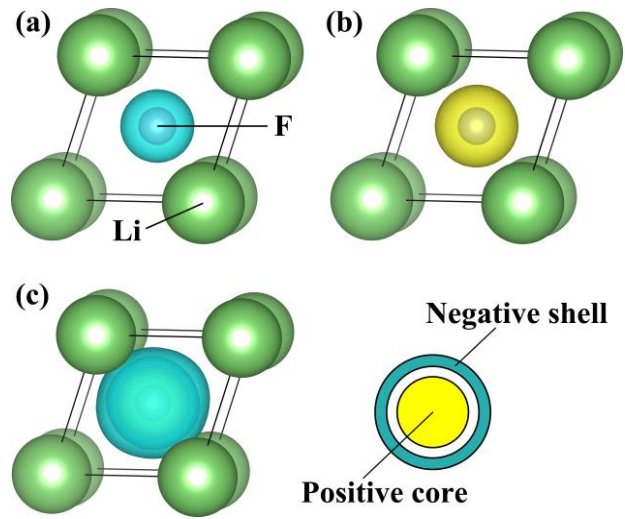

**Figure 4.** Charge density difference for a LiF primitive cell calculated using various methods. (a)

LDA compared with HSE06 ($\rho^{LDA}$ - $\rho^{HSE}$), where the cyan color means the difference is negative; (b) SC LDA-1/2 compared with HSE06 ($\rho^{SC-LDA-1/2}$ - $\rho^{HSE}$), where the yellow color stands for a positive difference; (c) $\rho^{NSC-LDA-1/2}$ - $\rho^{SC-LDA-1/2}$, where yellow and cyan colors stand for positive and negative differences, respectively. The absolute value of contour density is always set to 0.04 Å$^{-3}$.

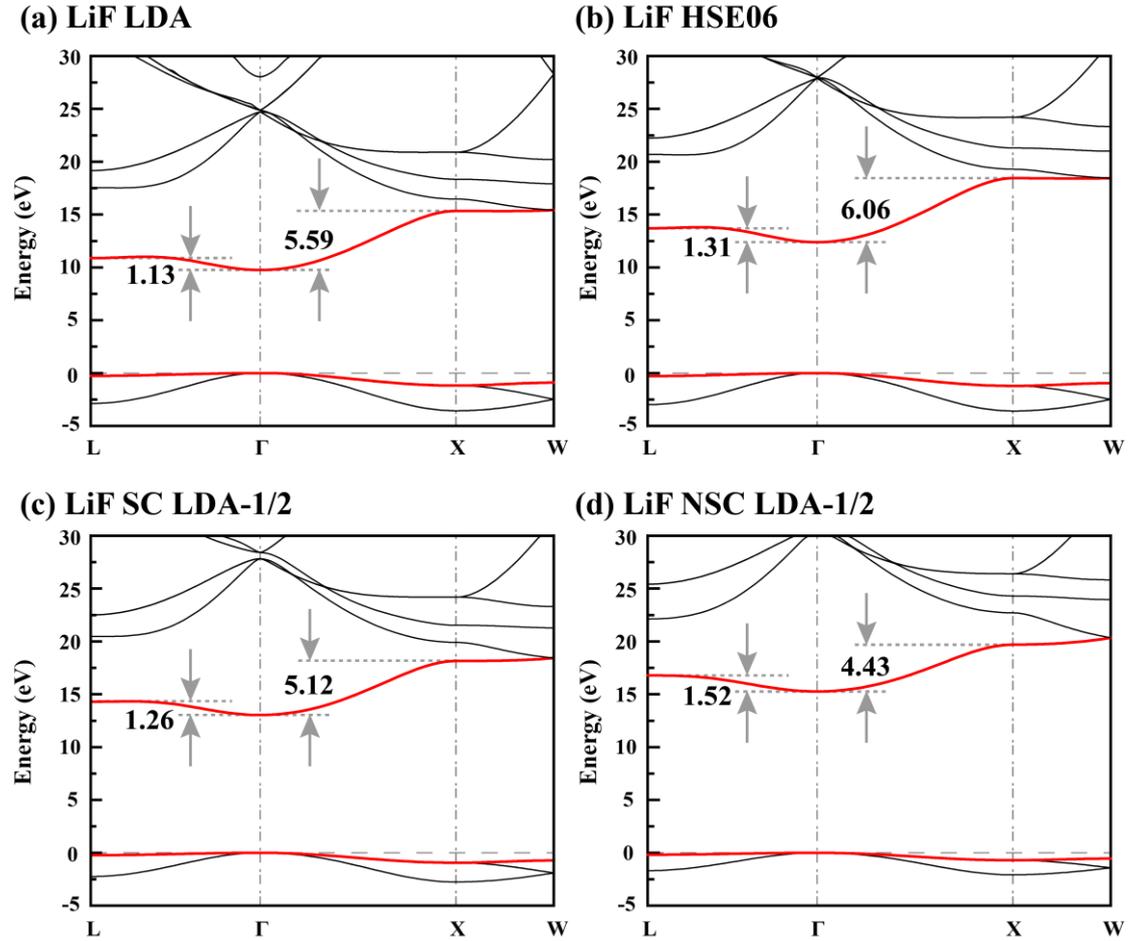

**Figure 5.** Electronic band structure of LiF, calculated using (a) standard LDA; (b) HSE06 hybrid functional based on the LDA-optimized lattice; (c) SC LDA-1/2, with optimized $r_{cut}$ = 2.1 bohr; (d) NSC LDA-1/2 based on the LDA ground state charge density, with optimized $r_{cut}$ = 1.8 bohr.

On the other hand, NSC LDA-1/2 calculations on LiF yields a band gap of 15.03 eV ($r_{cut}$ = 1.8 bohr), which at first glance seems very close to experimental. **Figures 4(c)** shows that in this case there are more electrons entering the SEP region, compared with SC LDA-1/2. These electrons originate mainly from an outer shell uncovered by the SEP. This is reasonable because the inter-electron repulsion energy inside the SEP region is counted only at the level of $\rho^{LDA}$, thus the densely populated electron densities within the SEP regions do not contribute much positive repulsion energy to the total energy. **Figures 5(a)-5(d)** illustrate the detailed electronic energy band structures of LiF calculated using LDA, HSE06, SC LDA-1/2 as well as NSC LDA-1/2. The topology of the CB is relative consistent among LDA, HSE06 and SC LDA-1/2 calculations. However, compared with the HSE06 results, in NSC LDA-1/2 the CB is too high at L with respect to Γ (+1.52 eV versus +1.31 eV), but is too low at X with respect to Γ (+4.43 versus with +6.60 eV). Meanwhile, the stress is enormously high, ~258.7 GPa along each direction in NSC LDA-1/2. The distorted CB is probably related to the lack of relaxation and the high stress level. While the fundamental gap from NSC LDA-1/2 calculation fits experimental for LiF, it is inferred that the overall NSC LDA-1/2 band structure suffers from abnormal distortion.

A similar trend is observed in CsCl, but the main distortion in its NSC LDA-1/2 band structure occurs in the VB rather than in the CB (**Figure 6**). From partial density of states analysis, it is observed that the VBs corresponds to Cl *p*-states, but the Cs states are just below. In HSE06 calculations, the separation between these two sets of bands is ~4 eV. Nevertheless, in SC DFT-1/2 the separation is less than 2 eV. Moreover, for NSC DFT-1/2 the separation is even reduced to ~0.5 eV. Such discrepancy reflects some intrinsic inaccuracy of the DFT-1/2 (as well as shell DFT-1/2) method, that the band gap maximization lets the top of VB suffers from the greatest downshift, which tends to compress the VB width, as well as the distance between the VB and core states. The former is a well-known effect already pointed out by Ferreira and coworkers [25], though whether such compression is closer to experimental still deserved more in-depth analysis. The latter, however, is a negative impact brought by DFT-1/2. For LiF, the core states are very low in energy, so that such narrowing of inner gaps between band sets does not influence the electronic states close to the Fermi level. Nevertheless, the Cs semi-core states lie close to the true VB in CsCl, which imposes artificially strong band repulsion between Cs and Cl states. Note that this problem is even more severe for NSC DFT-1/2. Hence, even for ionic insulators, SC DFT-1/2 should be preferred to NSC DFT-1/2 in order to achieve more accurate band structure details.

On the other hand, SC DFT-1/2 should perform better than NSC DFT-1/2 because the Cl states are permitted to respond to the external SEP, getting raised in energy to some extent through the radial dipole. Indeed, such a relaxation effect is still insufficient for CsCl. Some authors suggested the CB correction [45], which has been applied to two-dimensional semiconductors. For CsCl this is apparently not a suitable choice. The CB correction is related to attaching an additional positive potential energy to the Cs cations, which ought to raise the levels for Cs-related bands, further narrowing the gap between Cl and Cs band sets.

It now becomes important to understand the origin of band gap underestimation in the SC LDA-1/2 calculation of highly ionic insulators such as LiF and CsCl. The potential energy of an electron has two origins, one from external potential, including core potentials plus possible SEPs, and the other from interaction with other electrons. When comparing SC LDA-1/2 with standard LDA, the SEP regions surrounding the anions impose extra negative potential energy for the electrons close to the anions. Therefore, there is a net electron transfer from outside into the SEP regions, which establishes a radial dipole whose direction is from the anion core to the outside. Since the charge density is permitted to vary, reaching self-consistency between the Hamiltonian and the wavefunctions, the consequence of such radial dipole is to alleviate the electron transfer and to hinder the downshift of VB. In NSC LDA-1/2, however, the radial dipole does not influence the Hamiltonian, as the Coulomb interaction part of the Hamiltonian is always consistent with the LDA ground state, thus NSC LDA-1/2 should predict a larger band gap than SC LDA-1/2 (*cf.* **Figure 1**).

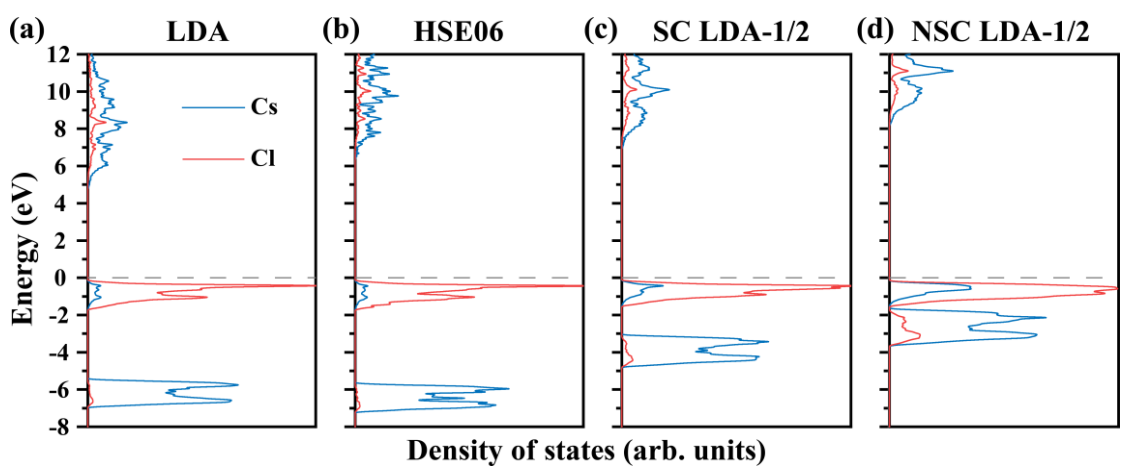

**Figure 6.** Projected electronic density of states of CsCl, calculated using (a) standard LDA; (b) HSE06 hybrid functional; (c) SC LDA-1/2; and (d) NSC LDA-1/2 based on the LDA ground state charge density.

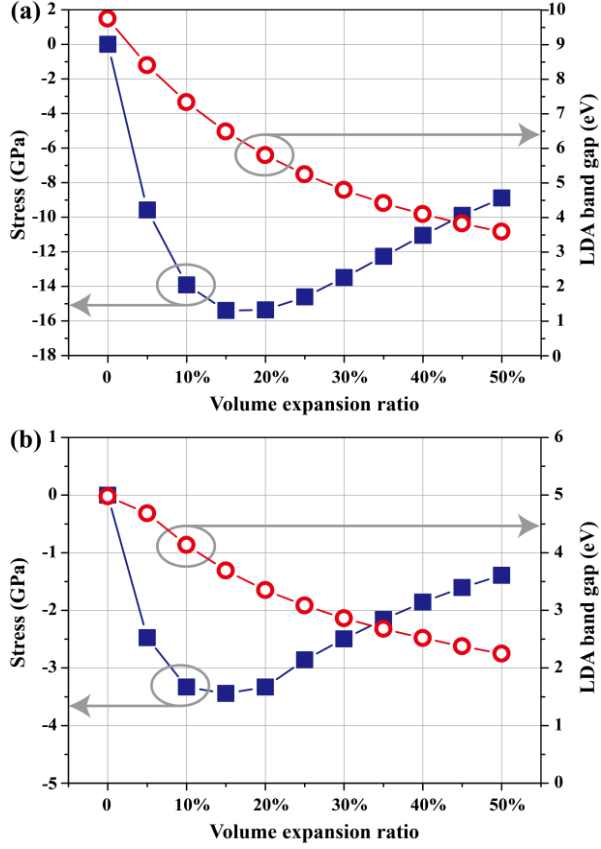

**Figure 7**. Test of stress levels along each direction as well as the LDA band gap values at various volume expansion ratios for (a) LiF primitive cell; (b) CsCl primitive cell.

Besides the radial dipole, it is also well-known that the band gap of a semiconductor or an insulator is sensitive to the lattice constant. Recently, we show that the band gap of LiF decreases monotonously upon enlarging the lattice [29]. The gap value is substantially lower when the stress is highly negative. However, the stress may not vary linearly with respect to the cell volume, thus we plot the dependence of stress as well as LDA band gap on volume expansion ratio, for LiF and CsCl, in **Figure 7**. Here the stress simply means the diagonal part of the stress tensor, and the three diagonal elements are actually equal due to the cubic symmetry. Interestingly, the magnitude of the stress does not increase all the way when the cell expands, for either material. Too severe lattice expansion may change the type of bonding and the electronic system has to respond to the new lattice constant. However, for the reasonable range of expansion, the negative stress indeed becomes more severe and the band gap decreases monotonously, when the cell volume increases from the equilibrium case. In SC LDA-1/2 calculation for LiF, it is observed that the stress is -29.8 GPa along each direction, thus the mechanical stress explains the low band gap value from SC LDA-1/2 calculation. This level of stress is in fact greater than any of the LDA value in **Figure 7(a)**. This is understandable because the external SEP does not alter the fundamental type of bonding in LiF and the lattice constant is still kept at the equilibrium LDA value. Hence, there is scope for the stress to rise to a high level under the influence of the strong SEP.

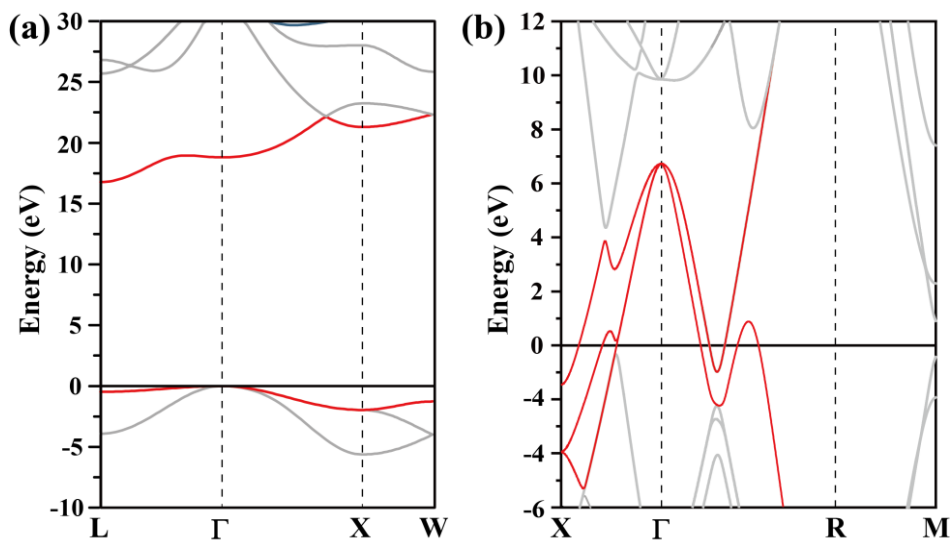

**Figure 8**. SC LDA-1/2 electronic energy band structures, calculated using lattice parameters optimized with SC LDA-1/2 corrected pseudopotentials: (a) LiF at 3.36 Å lattice constant; (b) CsCl at 2.41 Å lattice constant.

It is also interesting to relax the LiF primitive cell using the LDA-1/2-corrected F pseudopotential. A negative stress means that the LDA lattice constant is too large, viewed from the LDA-1/2-corrected pseudopotential perspective. The lattice constant is sharply reduced from 3.90 Å (LDA value) to 3.36 Å, which reaches a new equilibrium using the LDA-1/2 pseudopotentials. Therefore, based on the modified pseudopotentials, the LDA-predicted lattice constant is too large and should yield a smaller band gap due to the deformation potential [46]. Inversely, using 3.36 Å lattice constant, one expects a large band gap. Indeed, the band gap now increases to 16.77 eV, but the conduction band minimum (CBM) lies at the L point, rather than at Γ. **Figure 8(a)** indicates that the overall band structure is ill-shaped, even though the fundamental gap seems to have been enhanced against SC LDA-1/2 at LDA lattice constant. It follows that, at LDA level the most reasonable band structure of LiF still comes from SC LDA-1/2 calculation, though the band gap underestimation is related to the radial dipole as well as the unavoidable negative stress level. For CsCl, the relaxation using the Cl pseudopotential corrected by SC LDA-1/2 leads to a sharply reduced lattice constant of 2.41 Å, in comparison to the LDA-optimized value 3.98 Å. The resulting SC LDA-1/2 band diagram as shown in **Figure 8(b)** surprisingly shows a metallic behavior, which is due to the extreme compression.

## IV. SELF-CONSISTENCY OF DFT-1/2 IN COVALENT SEMICONDUCTORS

Before dwelling into the generic discussion of NSC DFT-1/2 for covalent semiconductors, a comparison between SC shell LDA-1/2 and NSC shell LDA-1/2 band diagrams for Ge is first given in **Figures 9(a)-9(c)**, where the LDA band diagram is also included for comparison, and all three calculations were carried out using LDA-optimized lattice constant (at 5.644 Å). SC shell LDA-1/2 and NSC shell LDA-1/2 both predict the CBM at the L point, and the band gaps are 0.80 eV and 0.86 eV, respectively, differing by only ~0.053 eV. As stated by Ferreira *et al.* [26], self-consistency is required in DFT-1/2 calculations for covalent semiconductors especially when the hole is scattered into multiple bands. Therefore, it is surprising to observe that the discrepancy between SC shell LDA-1/2 and NSC shell LDA-1/2 is much weaker in a typical covalent semiconductor like Ge, compared with typical ionic insulators like LiF. A charge analysis further confirms that the wavefunction and the charge density of Ge are less inconsistent in its NSC shell LDA-1/2 results. On the one hand, **Figure 10(a)** illustrates the spatial distribution for $\rho^{SC-LDA-1/2} - \rho^{LDA}$, which is already relatively small when viewed with the same critical contour density (0.04 Å$^{-3}$) as in **Figure 4**. On the other hand, $\rho^{NSC-LDA-1/2} - \rho^{SC-LDA-1/2}$ is even more tiny, and can be recognized only using a very small critical contour density 0.005 Å$^{-3}$, shown in **Figure 10(b)**. Hence, in NSC LDA-1/2 for Ge, the charge density derived from the wavefunction is quite similar to the corresponding LDA charge density that was used to construct the Hamiltonian. This explains the small band gap mismatch between SC and NSC shell LDA-1/2 calculations for Ge. On the other hand, the stress levels are quite similar in Ge calculations, both being negative (-39.28 GPa for SC shell LDA-1/2 and -38.90 GPa for NSC shell LDA-1/2 using the same cutoff function $r_{in}$ = 1.5 bohr and $r_{out}$ = 3.3 bohr). We also optimized the Ge cutoff radii specially for NSC shell LDA-1/2 calculation, but the optimized values remain the same ($r_{in}$ = 1.5 bohr and $r_{out}$ = 3.3 bohr).

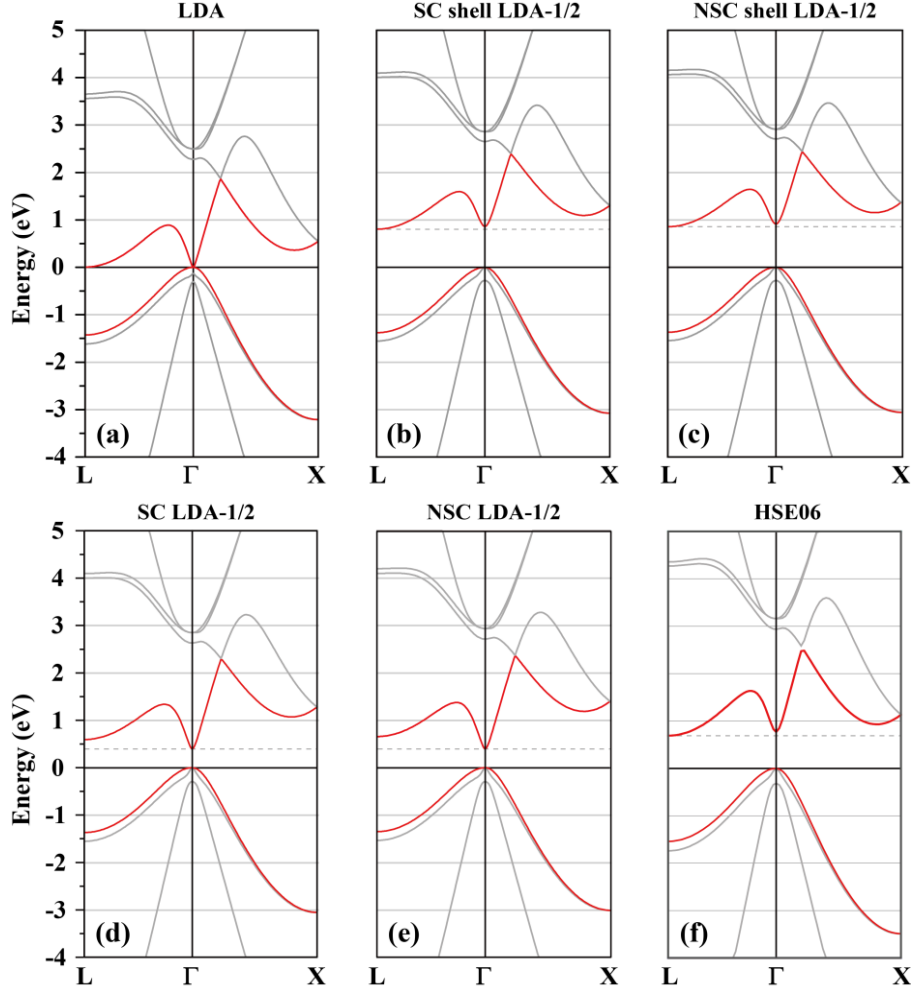

**Figure 9**. Electronic band diagrams of Ge, calculated using (a) LDA; (b) SC shell LDA-1/2; (c) NSC shell LDA-1/2; (d) conventional SC LDA-1/2; (e) conventional NSC LDA-1/2; (f) HSE06 hybrid functional. All calculations were carried out using LDA-optimized lattice constant 5.644 Å, and the level of CBM is indicated by a horizontal dashed line in each case.

An interesting issue is regarding the conventional LDA-1/2 calculation for Ge. Without an inner cutoff radius, the SC and NSC LDA-1/2 band diagrams for Ge are presented in **Figures 9(d)** and **9(e)**. In any case, a direct Γ–to–Γ band gap is revealed, inconsistent with well-known experimental evidence. Note that the stress levels of SC LDA-1/2 (-46.1 GPa at optimized $r_{cut}$ = 3.7 bohr) and NSC LDA-1/2 (-42.8 GPa at optimized $r_{cut}$ = 3.7 bohr, the same as in SC LDA-1/2) results are both higher than the corresponding shell LDA-1/2 cases. In a recent work, it was demonstrated that tensile stress tends to lower the CB of Ge at Γ, but raise the CB energy at L [29]. Shell LDA-1/2 avoids imposing too high stress levels through skipping an inner sphere region so as to shrink the volume for self-energy correction. It is also interesting to note that the HSE06 hybrid functional also reveals an indirect band gap feature of Ge, provided that the lattice constant is selected as the LDA-optimized one (**Figure 9(f)**).

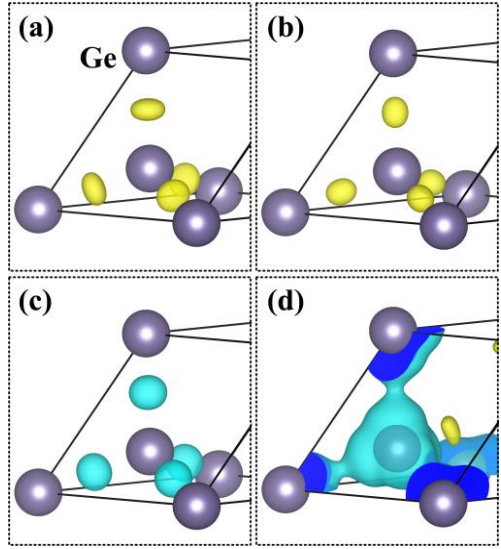

**Figure 10**. Electronic band diagrams of Ge, calculated using (a) $\rho^{SC\text{-}LDA\text{-}1/2} - \rho^{LDA}$, with contour density set to 0.04 Å$^{-3}$; (b) $\rho^{NSC\text{-}LDA\text{-}1/2} - \rho^{SC\text{-}LDA\text{-}1/2}$, with contour density set to 0.005 Å$^{-3}$; (c) $\rho^{LDA} - \rho^{HSE}$, with contour density set to -0.04 Å$^{-3}$; (d) $\rho^{SC\text{-}LDA\text{-}1/2} - \rho^{HSE}$, with contour density set to $\pm 0.005$ Å$^{-3}$. Yellow and cyan colors represent positive and negative values, respectively.

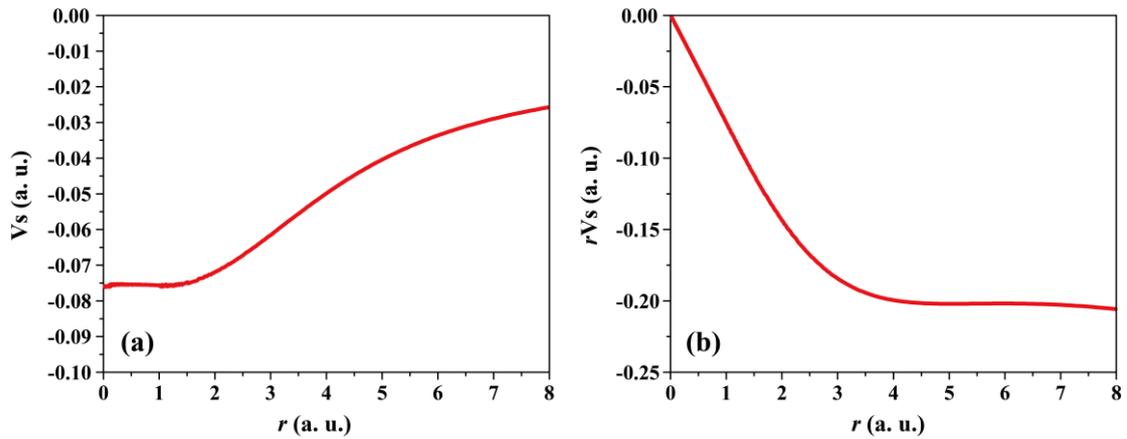

**Figure 11**. (a) The -1/4 e self energy potential ($V_S$) of Ge plotted against the radius $r$; (b) Plot of $rV_S$ versus $r$, which should converge to -1/4 at large $r$ limit, according to Gauss' law.

Charge comparison between LDA, HSE06 and SC LDA-1/2 calculations, all carried out using the LDA-optimized lattice, is presented in **Figures 10(c)** and **10(d)**. As expected, the LDA ground state suffers from the delocalization error as the VB electrons are less localized around the Ge-Ge bond centers than the HSE06 result (**Figure 10(c)**). On the other hand, the discrepancy between $\rho^{SC-LDA-1/2}$ and $\rho^{HSE}$ is very minor, and in **Figure 10(d)** we have to use small contour densities $\pm 0.005$ Å$^{-3}$ for illustration. Strong overcorrection like the LiF case is not observed, which is understandable because the SEPs are relatively weak far from the Ge cores, where the holes are localized. This is evidenced by the Ge SEP plot shown in **Figure 11**. Hence, both the strength of self-energy correction and the radial dipoles are relatively weak in Ge compared with highly ionic insulators.

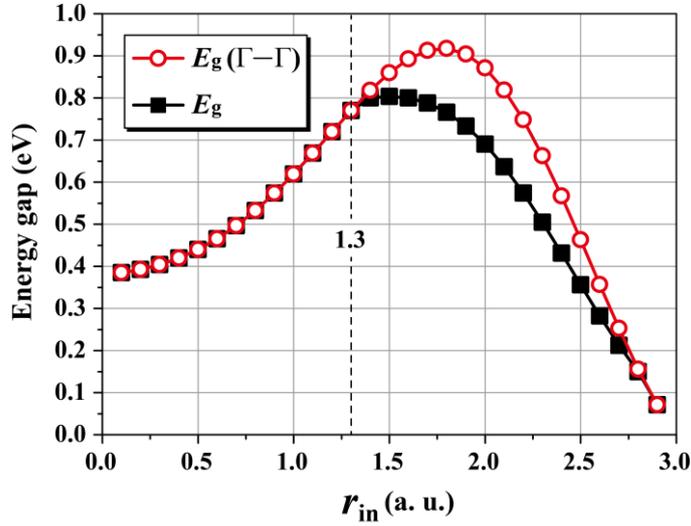

**Figure 12**. The shell LDA-1/2 energy gaps for Ge calculated at various $r_{in}$ values, fixing $r_{out}$ = 3.3 bohr. Both the fundamental gaps and the direct Γ-to-Γ gaps are shown.

We next compare the stress in SC and NSC shell LDA-1/2 for Ge. For more intuitionistic comparison, we fix the outer cutoff radius $r_{out}$ (3.3 bohr), but gradually decrease $r_{in}$ to convert shell LDA-1/2 into LDA-1/2. **Figure 12** illustrates the variations of the fundamental gap $E_g$ as well as the direct Γ-to- Γ gap with respect to $r_{in}$, showing clearly that $E_g(\Gamma-\Gamma)-E_g(\Gamma-L)$ switches its sign at an intermediate $r_{in}$ (1.3 bohr). Ge is predicted to be a direct gap semiconductor for $r_{in}$ = 1.3 bohr ($r_{out}$ still kept at 3.3 bohr). The stress is also related to $r_{in}$, rising from –39.3 GPa at $r_{in}$ = 1.5 bohr to -40.8 GPa at $r_{in}$ = 1.3 bohr. The reduced stress level in shell LDA-1/2 ($r_{in}$ = 1.3 bohr; $r_{out}$ = 3.3 bohr) is somehow related to the rectified band structure of Ge, though the difference is minor since 1.3 bohr and 1.5 bohr are very close.

Another typical covalent semiconductor under investigation is InP. It is a technically important semiconductor, which has been widely used in optical communications [47]. Obviously, InP has a mixed covalent-ionic bonding, but it is typically regarded as a covalent III-V semiconductor with a symbolic zinc blende structure. As pointed out in the 2018 work [28], shell LDA-1/2 for InP should be carried out in the shell LDA-1/4-1/4 manner (see also the **Appendix**). **Figure 13** illustrates its electronic band diagrams calculated using LDA, SC shell LDA-1/2 (In: $r_{in}$ = 2.3 bohr, $r_{out}$ = 4.2 bohr; P: $r_{in}$ = 1.4 bohr, $r_{out}$ = 3.3 bohr) as well as NSC shell LDA-1/2 (using the same set of cutoff radii). Like Ge, the NSC shell LDA-1/2 result is not very different from that of SC shell LDA-1/2, as the fundamental gap only increase mildly from 1.53 eV (SC shell LDA-1/2) to 1.77 eV (NSC shell LDA-1/2). In the mean time, the diagonal stress value changes very slightly from -40.64 GPa to -37.98 GPa, showing a consistent trend as Ge. However, the optimized cutoff radii for NSC LDA-1/2 are different, and the NSC LDA-1/2 gap is maximized to 2.11 eV with the SEP regions closer to P (In: $r_{in}$ = 2.6 bohr, $r_{out}$ = 5.9 bohr; P: $r_{in}$ = 0.7 bohr, $r_{out}$ = 3.1 bohr). The case of InP is distinct from Ge because there is additional freedom to adjust the four cutoff radii to reach a larger NSC band gap, thanks to the lack of self-consistency. The such-derived NSC LDA-1/2 band gap is even farther from experimental value 1.42 eV [48], greater by nearly 50%.

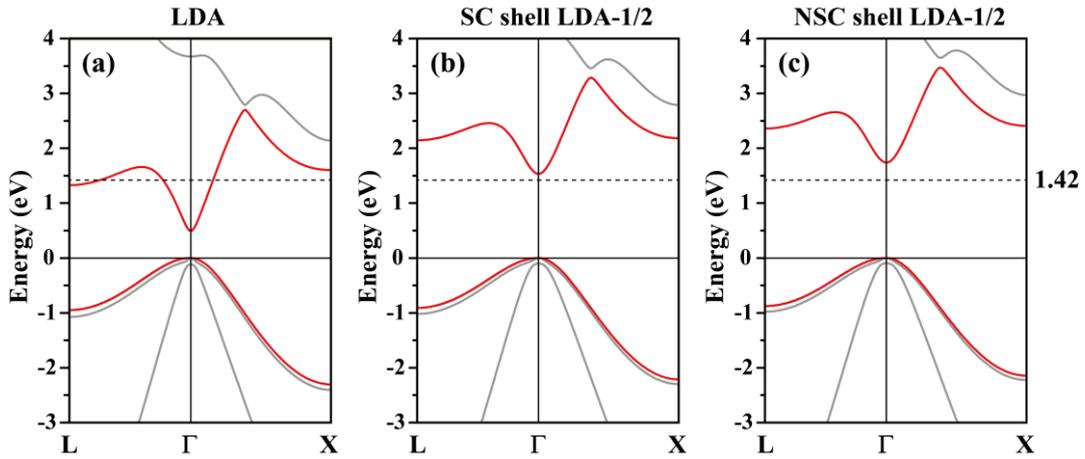

**Figure 13**. Electronic band diagrams of InP, calculated using (a) LDA; (b) SC shell LDA-1/2; (c) NSC shell LDA-1/2 with the same cutoff radii as in SC shell LDA-1/2. The tops of VBs are always set to zero energy, while the extrapolated zero temperature experimental gap (1.42 eV) is marked by dashed lines.

# V. INTERMEDIATE CASES

By intermediate cases we refer to those ionic insulators with yet a substantial proportion of covalent bonding. We select $TiO_2$ and $Ta_2O_5$ as typical examples. For $TiO_2$ the rutile phase is selected ($r$-$TiO_2$) for its simpler structure, while the newly discovered ground state γ-phase ($I4_1/amd$ symmetry) is used for $Ta_2O_5$. Their density of states analysis using LDA shows that the VB consists of states from both O and the metal element (**Figure 14**). A thorough SEP cutoff radii scan indicates that both materials require conventional LDA-1/2 calculations (*i.e.*, $r_{in}$=0). The SC LDA-1/2 band gap values (**Figures 15(a)** and **15(b)**) are quite satisfactory for both materials. The NSC LDA-1/2 band structures (**Figures 15(c)** and **15(d)**), however, show surprising results with severe gap overestimation. The differential charge analysis (illustrated in **Figure 16**, taking $r$-$TiO_2$ as an example), cutoff radii comparison, as well as Bader charge analysis all reveal that NSC LDA-1/2 renders $r$-$TiO_2$ and γ-$Ta_2O_5$ more ionic than they ought to be. For instance, NSC LDA-1/2 calculations drive more electrons towards the near-core region of O in $r$-$TiO_2$, and the source of these electrons is the Ti cation regions, as verified by the cyan parts in **Figure 16(c)**. If $r_{cut}$ is optimized for O in NSC LDA-1/2, it decreases from 2.4 bohr as in the SC LDA-1/2 case to 2.0 bohr, preferring a narrower SEP correction region, which is typical for highly ionic compounds as well (see **Sect. III**). In γ-$Ta_2O_5$, it is also observed that the optimized O $r_{cut}$ for NSC LDA-1/2 (1.9 bohr) is smaller than the optimized $r_{cut}$ for SC LDA-1/2 (2.3 bohr).

Bader charge results with the same $r_{cut}$ reveals that the amount of positive charge on Ti/Ta cations increases from +2.64 e/+3.43 e, as in SC LDA-1/2, to +3.32 e/+4.12 e as in NSC LDA-1/2. The Bader analysis for NSC LDA-1/2 is carried out on the new charge density derived from the electronic wavefunctions (otherwise it is the original LDA charge density that is analyzed, which yields merely +2.21 e/+2.94 e for Ti in $TiO_2$ and Ta in $Ta_2O_5$, respectively). In contrast, the Bader charge of Li in LiF only slightly increases from +0.92 e in SC LDA-1/2 calculation to +0.94 e in NSC LDA-1/2. The greatest mismatch between SC and NSC LDA-1/2 band gaps occurs in intermediate cases with mixed ionic-covalent bonding, which at first glance seems peculiar. Nevertheless, a proper explanation is easily reached considering the enhancement of ionicity brought about by NSC LDA-1/2. On the one hand, highly ionic insulators such as LiF and CsCl have little scope to become even more ionic. On the other hand, the stability of covalent bonds, stemming from saturation and directional properties, also prohibits very strong deviation from the original bonding characteristics. The exceptional case is a generally ionic insulator with a certain degree of covalent bonding. In NSC DFT-1/2 calculations, the degree of covalent bonding is suppressed and ionicity is enhanced, rendering the observed abnormally large band gaps. Of course, the enhanced ionicity is an artifact due to the absence of self-consistency between the wavefunction and the charge density involved in the Hamiltonian.

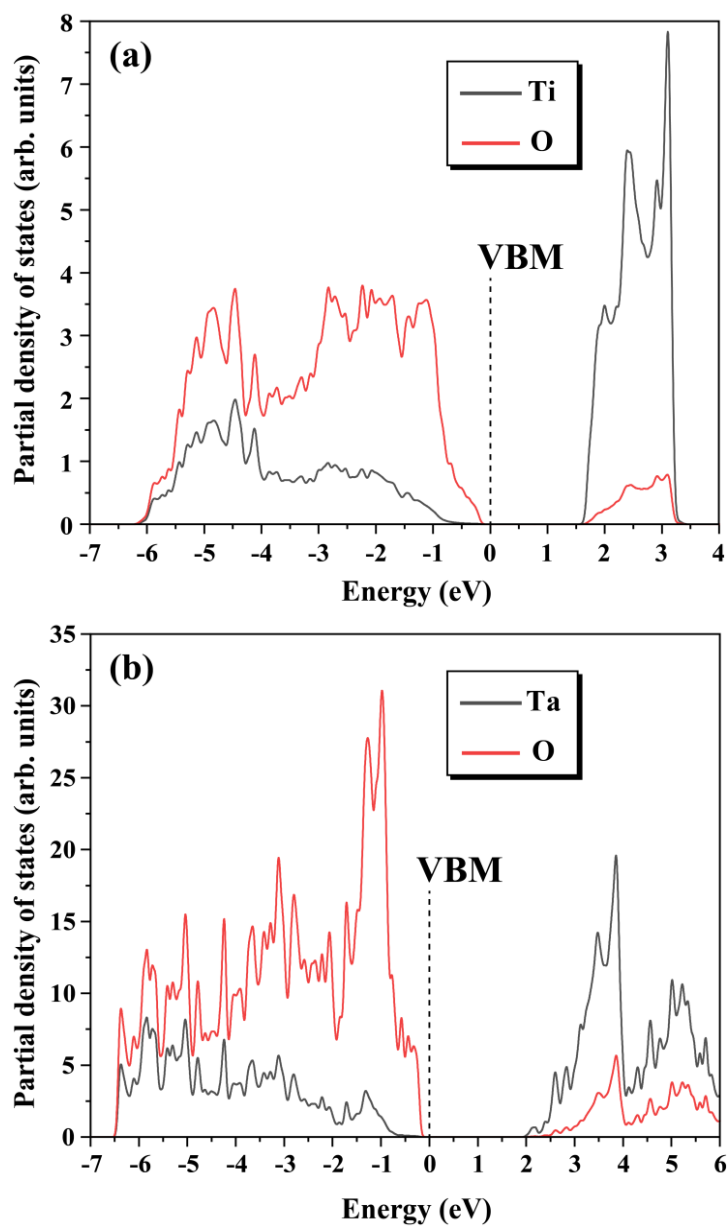

**Figure 14**. Partial density of states of (a) rutile $TiO_2$ and (b) $\gamma$-$Ta_2O_5$, calculated using LDA.

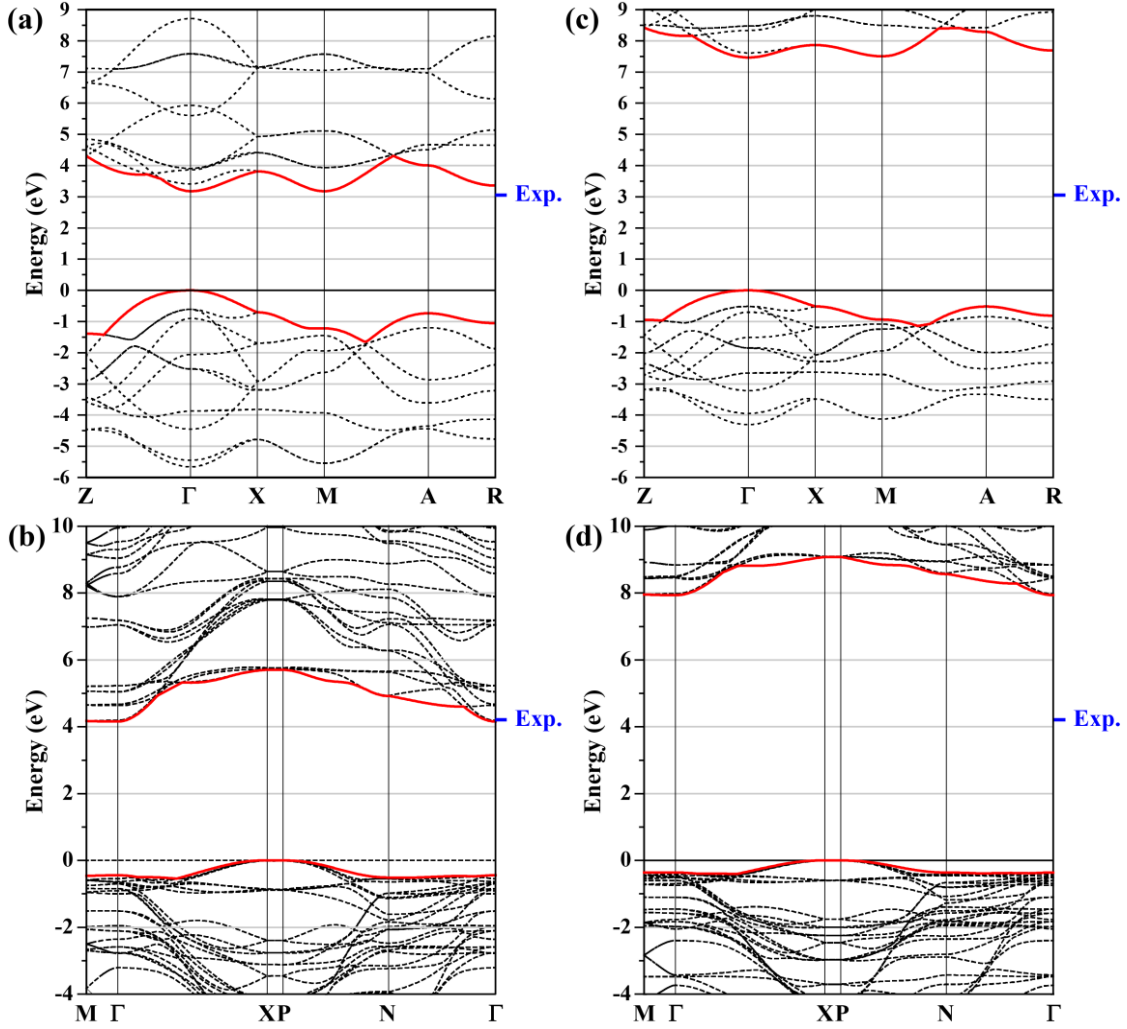

**Figure 15**. Electronic energy band diagrams of (a) rutile TiO$_2$ calculated with SC LDA-1/2; (b) γ-Ta$_2$O$_5$ calculated with SC LDA-1/2; (c) rutile TiO$_2$ calculated with NSC LDA-1/2; (d) γ-Ta$_2$O$_5$ calculated with NSC LDA-1/2. Experimental gaps: 3.05 eV [49] for single crystal rutile TiO$_2$; 4.20 eV [50] for Ta$_2$O$_5$.

## VI. ADDITIONAL DISCUSSION

For all compounds under investigation, it is observed that SC LDA-1/2 leads to negative stresses, as long as the lattice structure is kept the same as that optimized using LDA. The SEP regions are surrounding the anions, which tend to draw electrons closer to the anions, or to the bond center locations in terms of element covalent semiconductors such as Ge. The ions, on the other hand, are kept unmoved compared with the LDA lattice. This generates a tendency to draw ions closer to each other, and consequently the negative stress. Such stress level has a different origin from that stemming from lattice constant mismatch, and can be enormous in SC LDA-1/2. It is an artifact in DFT-1/2 because of the self-energy correction in real space, and it indeed causes certain inaccuracy regarding the band gaps, especially for highly ionic insulators. Besides the radial dipole that reduces the band gap, the negative stress is another reason that leads to band gap underestimation in SC LDA-1/2. Fortunately, the overall gap underestimation even in the extreme case of LiF, is still limited to ~1.1 eV provided that $p = 20$ is adopted as the power index in the cutoff function. Using $p = 8$ will add to the inaccuracy.

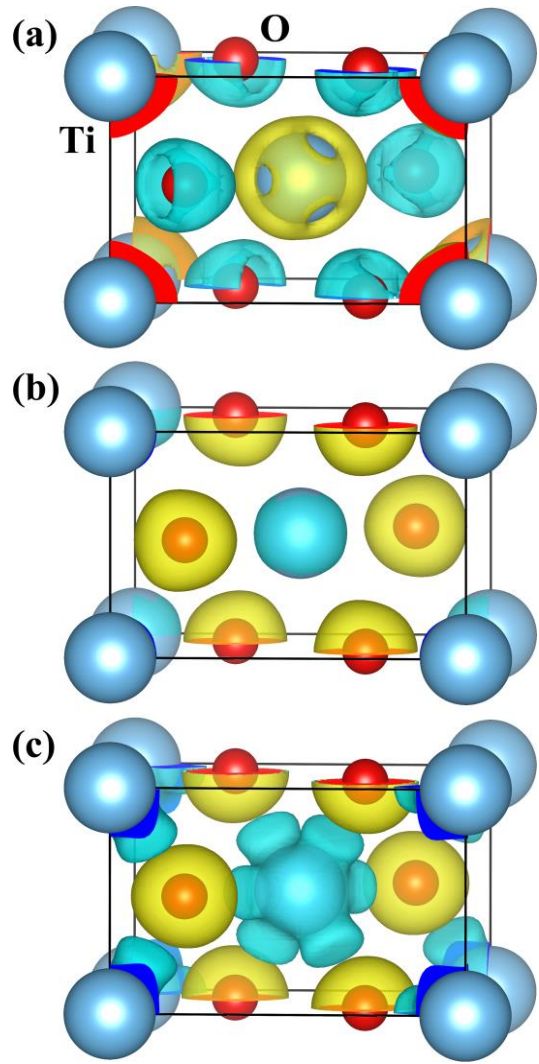

**Figure 16**. Differential charge analysis for rutile $TiO_2$. (a) $\rho^{LDA} - \rho^{HSE}$, with contour density 0.04 Å$^{-3}$; (b) $\rho^{SC-LDA-1/2} - \rho^{HSE}$, with contour density 0.04 Å$^{-3}$; (c) $\rho^{NSC-LDA-1/2} - \rho^{SC-LDA-1/2}$, with contour density 0.08 Å$^{-3}$. Yellow and cyan colors represent positive and negative values, respectively.

The negative stress is more severe if the SEP region is large, thus shell LDA-1/2 can alleviate the stress problem through shrinking the SEP region. For covalent semiconductor Ge, the finite inner radius not only increases the fundamental gap, but also inverses the relative positive at L and at Γ for the CB. In other words, the CB of Ge is artificially distorted by the SEP coverage at the near-core regions of Ge.

In NSC LDA-1/2 calculations, the stress goes towards the positive direction compared with SC LDA-1/2. For rutile $TiO_2$, the NSC LDA-1/2 stresses become close to zero. For the more ionic LiF, the NSC LDA-1/2 stress increases to an enormously high positive level (+258.7 GPa). Although the stress values for NSC calculations are questionable, this reflects the fact that electrons are densely packed in the SEP regions for NSC LDA-1/2, as reflected in their wavefunctions. The amount of transferred electrons is more than in SC LDA-1/2, because there is no penalty of counting their strong repulsion due to a fixed Hamiltonian.

## VII. CONCLUSION

Using LDA for the exchange-correlation, we have discussed the differences between SC DFT-1/2 and NSC DFT-1/2 calculations. The following conclusions have been drawn.

(i) The charge density distribution of SC LDA-1/2 is distinct from that of LDA, that electrons are more localized around the anions. The direction of correction is reasonable because LDA suffers from the delocalization error. However, SC LDA-1/2 overcorrects this to a great extent, because the external potential has changed from LDA. The Hohenberg-Kohn theorem guarantees that a different ground state charge density is derived for SC LDA-1/2.

(ii) NSC LDA-1/2 does not mean the charge density strictly follows that of LDA, because the LDA charge density is only used to construct the Hamiltonian. According to the electron wavefunctions of NSC LDA-1/2, there are much more electrons transferred to the regions covered by the self-energy potential, compared with SC LDA-1/2, as no extra repulsion energy is counted in NSC LDA-1/2 calculations. Although the NSC LDA-1/2 band gap may be closer to experimental for several highly ionic insulators, the detailed band structures usually suffer from distortions, which are inferior to that calculated by SC LDA-

1/2.

(iii)   SC LDA-1/2 generates radial dipoles due to the screening for the introduced self-energy potential, which tends to reduce the extent of band gap correction, and leads to slightly underestimated band gaps for highly ionic insulators. Moreover, SC LDA-1/2 uniformly causes negative (tensile) stresses in all the compounds investigated in this work. Such stress level can be much stronger than the conventional stress caused by lattice constant change. This is a special stress belonging to LDA-1/2 and can also reduce the band gap, since most insulators show reduced gaps under tensile stress.

(iv)   Compared with SC LDA-1/2 (or shell LDA-1/2 if the compound has covalent bonding), NSC LDA-1/2 (NSC shell LDA-1/2) leads to the greatest errors in some mixed ionic-covalent insulators/semiconductors exemplified by $TiO_2$. The reason lies in that NSC LDA-1/2 tends to increase the ionicity of the bonding. On the one hand, for purely ionic compounds, there is little scope for further enhancement of ionicity. On the other hand, purely covalent semiconductors can hardly be transformed to ionic due to the robust saturation nature of covalent bonds and the symmetry in elemental semiconductors. For semiconductors like $TiO_2$, they possess intermediate ionicity that may be enhanced substantially. Consequently, NSC LDA-1/2 tends to greatly increase their ionicity and leads to abnormally large band gaps.

To sum up, NSC LDA-1/2 has potentially uncontrollable influence to the conduction band and the detailed band structures for insulators and semiconductors. The enhanced ionicity in NSC LDA-1/2 calculations may also render abnormally large band gaps. It is suggested that SC LDA-1/2 should be globally implemented instead of NSC LDA-1/2 for ionic insulators, covalent semiconductors as well as mixed ionic-covalent compounds.

## Appendix. Computational settings

For all LDA, LDA-1/2 and shell LDA-1/2 calculations (regardless of self-consistent or non-self-consistent), the plane wave kinetic energy cutoff was constantly fixed to 600 eV. The Monkhorst-Pack $k$-point grid [51] for Brillouin zone sampling during geometry optimization and ground state charge calculations was 25 × 25 × 25 for those zinc blende structure semiconductors, centered at the Γ point. For other compounds, equal-spacing $k$-meshes with similar densities were adopted. Spin-orbit coupling was considered in Ge and InP calculations. The electrons considered as in the valence were: 1s and 2s for Li; 2s and 2p for O, F; 2s, 2p and 3s for Na; 3s and 3p for P, Cl; 3s, 3p and 4s for K; 4s and 4p for Ge; 4s, 4p and 5s for Rb; 5s and 5p for In; 3s, 3p, 3d, and 4s for Ti; 5p, 5d, and 6s for Ta; 5s, 5p and 6s for Cs.

The HSE calculations were based on the pseudopotentials originally derived with the Perdew-Burke-Ernzerhof functional [5]. The range separation parameter was $\omega = 0.106$ bohr$^{-1}$, a default value for HSE06 in VASP, and 25% Hartree-Fock exchange was mixed for the short range part. Spin-orbit coupling was turned on only for Ge. As Ge is a rather soft element, and the non-collinear calculations would greatly increase the number of irreducible $k$ points, we here, solely for Ge, used a small plane-wave basis with a cutoff energy of 350 eV for better efficiency. The HSE band structures of Ge were derived self-consistently using an equal-spacing $k$-mesh with finite weights, plus the particular $k$ points along the L-Γ-X lines in the Brillouin zone whose weights were yet set to zero.

DFT-1/2 self-energy corrections were carried out in the following manner: -1/2 e correction (or simply -1/2) for F, Cl, Br and I, regarding all their compounds involved in this work; -1/4 for Ge; -1/2 for O in $TiO_2$ or $Ta_2O_5$. Shell DFT-1/2 self-energy corrections were carried out in the following manner: -1/4 for Ge; -1/4 for In and -1/4 for P in InP; -1/2 for O in $TiO_2$ or $Ta_2O_5$. The self-energy potentials were derived using a modified ATOM code [52,53]. The power index $p$ [27] in the cutoff function of DFT-1/2 (including shell DFT-1/2) is constantly chosen as 20 [28].

## Acknowledgement

This work was supported by the National Natural Science Foundation of China under Grant No. 61974049.